\documentclass{kluwer}
\usepackage[]{graphicx}
\begin{document}

\def\aj{{\it AJ}}
\def\apj{{\it ApJ}}
\def\apjs{{\it ApJS}}
\def\aap{{\it A\&A}}
\def\pasp{{\it PASP}}
\def\mnras{{\it MNRAS}}
\def\teff{$T_{\rm eff}$}
\begin{article}
\begin{opening}
\title{On ionisation effects and abundance ratios in damped Lyman-$\alpha$ systems}
      

\author{D. \surname{Schaerer}\email{schaerer@ast.obs-mip.fr}}
\institute{Observatoire Midi-Pyr\'en\'ees, F-31400 Toulouse, France}
\author{Y.I.. \surname{Izotov}\email{izotov@mao.kiev.ua}} 
\institute{Main Astronomical Observatory, Kiev 03680, Ukraine}
\author{C. \surname{Charbonnel}\email{corinne@ast.obs-mip.fr}}
\institute{Observatoire Midi-Pyr\'en\'ees, F-31400 Toulouse, France}


\runningtitle{Ionisation effects and abundance ratios in DLAs}
\runningauthor{Schaerer, Izotov, Charbonnel}

\begin{abstract} 
The similarity between observed velocity structures of Al III and singly ionised 
species in damped Ly$\alpha$ systems (DLAs) suggests the presence of ionised gas
in the regions where most metal absorption lines are formed.
To explore the possible implications of ionisation effects we construct
a simplified two-region model for DLAs consisting of an ionisation
bounded region with an internal radiation field and a neutral region with a 
lower metal content. Within this framework we find that ionisation effects 
are important.
If taken into account, the element abundance ratios in DLAs are quite
consistent with those observed in Milky Way stars and in metal-poor H II regions
in blue compact dwarf galaxies.
In particular we cannot exclude the same primary N origin in both DLAs and 
metal-poor galaxies. From our models no dust depletion of heavy elements needs 
to be invoked; little depletion is however not excluded.

\end{abstract}

\end{opening}

\vspace*{-0.5cm}
\section{Introduction}
The study of damped Lyman $\alpha$ systems (DLAs) is of great interest
for numerous fields in astrophysics (cf.\ Petitjean, these proceedings), 
including nucleosynthesis and the chemical evolution of galaxies and the 
Universe at large.
In addition to investigations on the metallicity evolution with redshift
(e.g.\ Pettini et al.\ 1994, 1997),
spectroscopic studies with 10m class telescopes have opened the 
opportunity to study Ly$\alpha$ absorption lines for a large number of 
elements including N, O, Mg, Al, Si, S, Cr, Mn, Fe, Ni and Zn.

Two main interpretations of the abundance ratios in DLAs exist currently
in the literature -- both suffering from several inconsistencies.
{\bf 1)} Lu et al.\ (1996), Prochaska \& Wolfe (1999), Outram et al.\ (1999), and
Pettini et al.\ (2000) found that the relative abundance patterns 
indicate that the bulk of heavy elements in these high-redshift objects
were produced by Type II supernovae.
However, when combined with the effects of dust depletion (as indicated by 
the relative overabundance of Zn relative to Cr or Fe) this interpretation is 
inconsistent with other elemental abundance ratios such as N/O, S/Fe, Mn/Fe and Ti/Fe. 
This led Lu et al.\ (1996) and Prochaska \& Wolfe (1999) to conclude that the 
overabundance of Zn/Cr in DLAs may be intrinsic to stellar nucleosynthesis in
such systems; a new physical process is then required to explain the
damped Ly$\alpha$ abundance patterns. 
{\bf 2)} Pettini et al.\ (1997, 1999), Vladilo (1998) and others argue  
that the relative abundances of Cr, Fe, and Ni are consistent with a moderate 
degree of dust depletion that, once accounted for, leaves
no room for the enhancement of $\alpha$ elements over iron
seen in metal-poor stars in the Milky Way. This is in contradiction with 
a SNII enrichment pattern,  but it could be understood if star formation
in these systems proceeded at a lower rate than in the early history
of our Galaxy (e.g.\ Pettini et al.\ 1999; Centuri\'on et al. 2000).
A similar conclusion is also drawn from the Fe/H vs. redshift distribution of
DLAs when compared with our Galaxy (e.g.\ Lu et al.\ 1996; Prochaska \& Wolfe 
2000).

It is generally believed that the gas in DLAs is mostly
in the neutral stage due to the very high optical depth beyond the
hydrogen ionisation limit. Therefore, it is assumed that
all species with ionisation potentials lower than hydrogen
are in a singly ionised stage, while other species are neutral.
However, Lu et al.\ (1996) and Prochaska \& Wolfe (1999) find a
good correlation between the velocity structure of Al III and
singly ionised species. The ionisation potential of Al$^+$
is 18.8 eV, i.e. greater than that of hydrogen. Therefore,
Al$^{+2}$ is likely present in ionised, not in neutral gas.
To explain the similarity of Al III and other low ionisation
species line profiles, Howk \& Sembach (1999) and Izotov \& Thuan (1999)
proposed that these lines originate in the same ionised region 
or in a mix of neutral and ionised clouds, and stressed the importance
of abundance correction for ionisation effects. 
In the present work we adopt a simple partially ionised two-region model 
for DLAs and explore the possible implications on abundance ratio
determinations within this framework. Full account of this work
is given in Izotov et al.\ (2000, hereafter ISC).

\vspace*{-0.5cm}
\section{A simplified model}
In the present paper we explore the following ``multi-component'' 
picture for DLA systems: 
{\em 1)} an ionisation bounded region illuminated by an internal 
radiation field complemented by {\em 2)} a neutral region with a 
lower (negligible) metal content.
Region 1 is modeled with the ionisation equilibrium code CLOUDY 
assuming internal stellar radiation and 
plane-parallel geometry. The outer boundary of the slab is determined
by the condition that a low electron temperature of 2000 K is reached.
We adopt stellar ionising spectra with various temperatures \teff, 
and a metal abundance of 
0.1 solar with relative element abundances equivalent to those observed 
in low-metallicity BCDs (e.g., Izotov \& Thuan 1999).
The ionisation parameter $U$ is varied over a wide range.
Constraints on $U$ are derived from the observed Al$^{2+}$/Si$^+$ 
column densities.
To explore the average effects, a single ``typical'' value 
($\log U \sim$ -3) is derived from the
typical ratio $\log$ [$N$(Al III)/$N$(Si II)] $\sim$ -2 to -2.5 
in DLAs. 
The corresponding ionisation correction factors, which are found
to be non-negligible, are applied to all observed abundance ratios 
to be discussed below (Fig.\ 2).

A similar model was recently considered by Howk \& Sembach (1999).
The main difference with their approach is that they consider 
ionised density-bounded regions surrounding the neutral, H I--bearing 
clouds. Both qualitatively and quantitatively the resulting ionisation 
corrections are quite different from our model. For more details see ISC.

An important feature of both models is that the heavy element abundances
in the ionised and neutral regions are a priori unrelated.
This implies in particular that relative abundance determinations
with respect to hydrogen (e.g. [Fe/H], [Zn/H]), as derived from the 
observed column densities, may not be meaningful or are at least uncertain.

\vspace*{-0.5cm}
\section{Implications on abundance ratios in DLAs}

Figure 1 shows that the observed correlation between [Si/Fe] 
and [Zn/Fe] which is attributed to progressive dust depletion 
(Prochaska \& Wolfe 1999) can also be explained by ionisation effects.
The best consistency is obtained for the radiation field corresponding 
to \teff=40 kK and $U$ values compatible with the observed Al III/Si II
column density ratios.
In this case one finds also that after the ionisation correction is applied
(and assuming a final value [Zn/Fe] = 0), the resulting [Si/Fe] values 
remain positive (reminiscent of $\alpha$-element enrichment). 
No additional correction for dust depletion is required, but little depletion
is not excluded.

\begin{figure}[htbp]
\centerline{\includegraphics[width=7.5cm,angle=270]{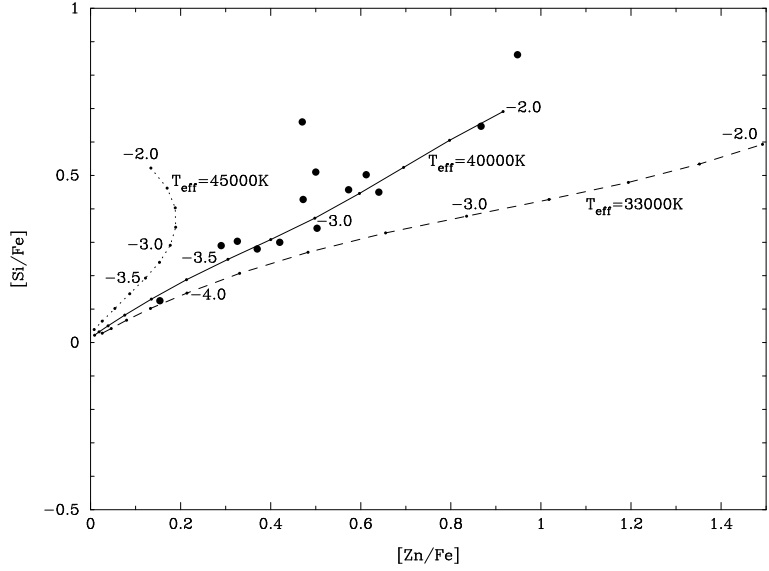}}
\caption{[Si/Fe] versus [Zn/Fe] showing observations from Prochaska \& 
Wolfe (1999) and model sequences for different values of \teff\ and
$U$.}
\end{figure}
\begin{figure}[htbp]
\centerline{\includegraphics[width=7.5cm,angle=270]{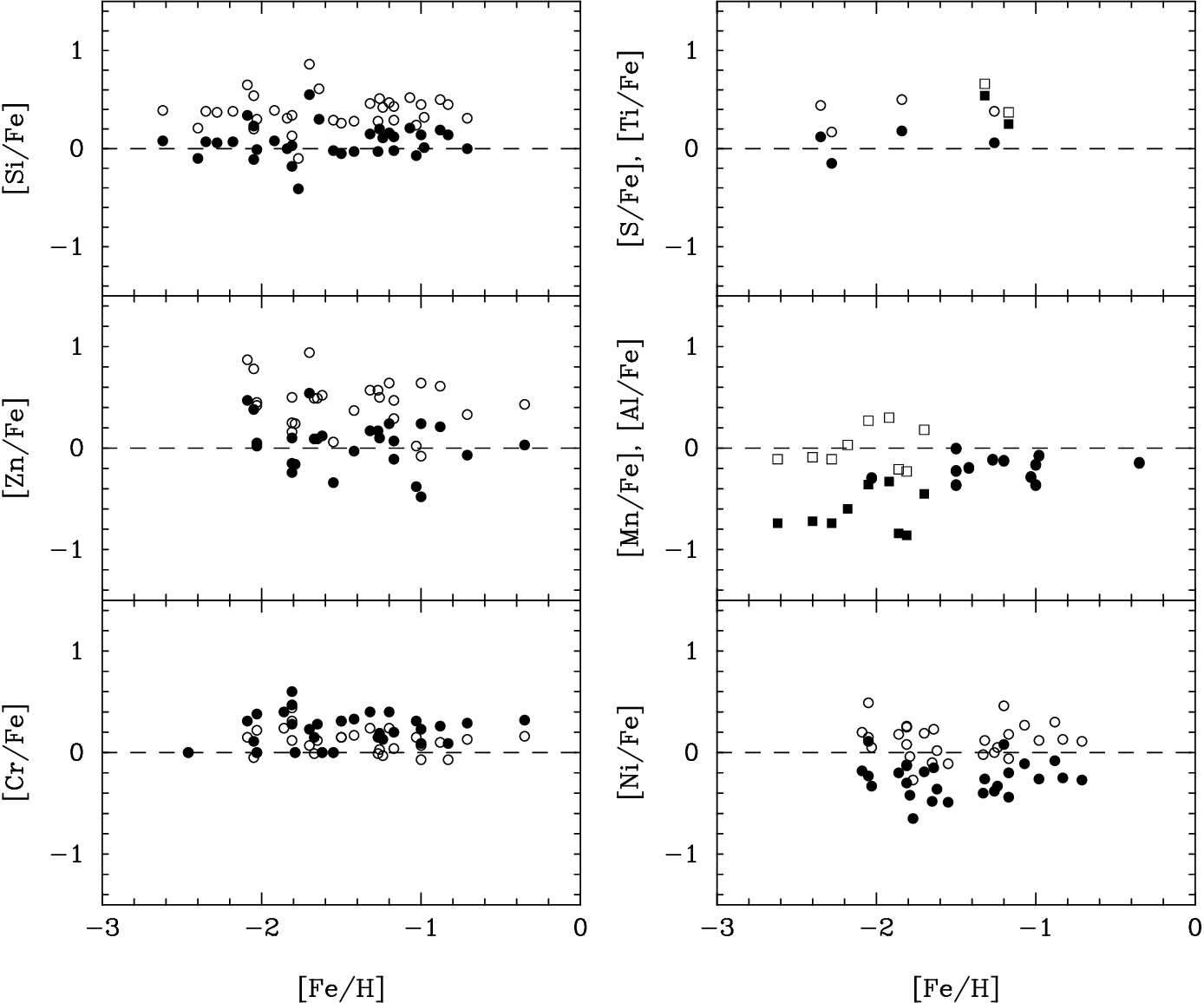}
            \includegraphics[width=6cm,angle=270]{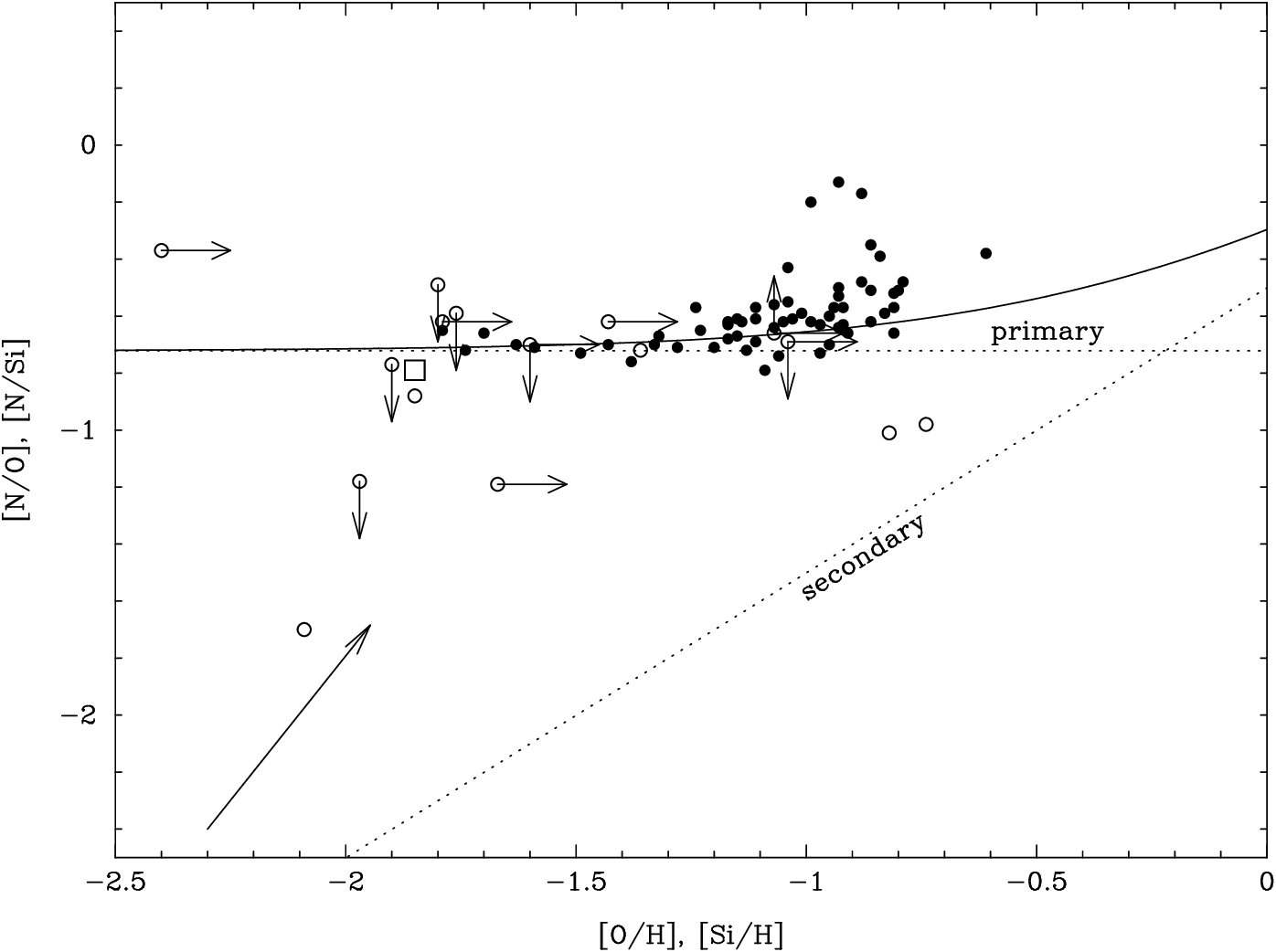}}
\caption{{\bf Left:} Corrected (filled symbols) and uncorrected (open) abundance 
ratios based on the data of Lu et al.\ (1996), Prochaska \& Wolfe (1999) 
and Pettini et al.\ (2000) as a function of [Fe/H] (uncorrected).
{\bf Right:} Observed N/O and N/Si ratios in BCDs (filled circles: Izotov \&
Thuan 1999) and DLAs (open circles: Lu et al.\ 1998, Outram et al.\ 1999)
as a function of metallicity. The typical ionisation correction is indicated
by the arrow.}
\end{figure}

The corrected and uncorrected abundance ratios of various elements
are shown in Figure 2 (left). Except for Cr/Fe, all element ratios with respect
to Fe have downward corrections with typical values between $\sim$ 0
(Mn/Fe), -0.4 (Zn/Fe) and -0.6 dex (Al/Fe). [With respect to Zn all ratios
except for Al/Zn have to be corrected upwards.]
After correction we find that S and Si (overabundant in [X/Fe]), Mn and Al 
(underabundant), Ti (small overabundance), and Zn ($\sim$ solar) follow quite 
well their respective SNII pattern observed in metal-poor halo stars.
Cr and Ni (including the oscillator strengths of Fedchak \& Lawler 1999)
show some disagreement whose origin is not understood yet (atomic data ?
other ?). See ISC for a more detailed discussion.

The right panel in Fig.\ 2 shows observed N/O and N/Si ratios from metal-poor
HII regions and measurements in DLAs respectively.
From our model we find a typical upward correction of N I/Si II by 0.7 dex,
as shown by the arrow. Once accounted for, it appears that the majority 
of measurements for DLAs are at or above the N/O value from the BCDs.
This suggests that a significant part of the difference previously found
in [N/Si] between HII regions and DLAs (Lu et al.\ 1998) can be 
explained by ionisation effects.
Interestingly, the recent direct measurement of N/O in a DLA from unsaturated 
OI and NI lines (open square; Molaro et al.\ 2000),
which are also little dependent on ionisation corrections,  
agrees as well with the metal-poor BCDs.
From our analysis we conclude that, if our typical ionisation corrections
apply, the present data shows no significant difference in N/O 
abundance ratios between low-metallicity BCDs and DLA systems implying
a similar origin of these two elements.

Implications on the metallicity-redshift relation of DLAs and other issues
are discussed in ISC.

A detailed discussion of our assumptions and support from
other data are presented in ISC. 
These working hypothesis have been adopted to study the main
effects intrinsic stellar ionising radiation may have on 
observed heavy element patterns in such conditions.
Future investigations should hopefully be able to improve on
these issues and assess the validity of our picture.
It is the hope that the proposed simplified model reflects the main
trends due to ionisation effects in DLAs. If correct, 
our picture offers a clear simplification in the understanding of 
heavy element abundance ratios in DLAs and their comparison with the 
local Universe.

\end{article}
\end{document}